\documentclass[aps,prl,twocolumn,preprintnumbers,amsmath,nofootinbib,longbibliography]{revtex4-1}

\usepackage{amsmath, amssymb,amsfonts,mathrsfs}
\usepackage{color}
\usepackage{extarrows}

\usepackage[colorlinks=true, linkcolor=blue, urlcolor=blue, citecolor=blue]{hyperref}

\newcommand\ud{\mathrm{d}}

\begin{document}
\preprint{{\tt CP3-18-56},\,{\tt MCnet-18-27}} 

\title{\Large\sffamily\bfseries
Bootstrapping solutions of scattering equations}

\author{{\sc Zhengwen Liu} and {\sc Xiaoran Zhao}}
\affiliation{Centre for Cosmology, Particle Physics and Phenomenology (CP3), UCLouvain,\,1348 Louvain-la-Neuve,\,Belgium}

\begin{abstract}
\noindent
The scattering equations are a set of algebraic equations connecting the kinematic space of massless particles and the moduli space of Riemann spheres with marked points.
We present an efficient method for solving the scattering equations based on the numerical algebraic geometry.
The cornerstone of our method is the concept of the physical homotopy between different points in the kinematic space, which naturally induces a homotopy of the scattering equations.
As a result, the solutions of the scattering equations with different points in the kinematic space can be tracked from each other.
Finally, with the help of soft limits, all solutions can be bootstrapped from the known solution for the four-particle scattering.
\end{abstract}

\maketitle

\vskip 5pt
\noindent{\sffamily\bfseries Introduction~}
The {\it scattering equations} are a set of algebraic equations connecting the kinematic space ${\cal K}_n$ of $n$ massless particles spanned by linearly independent Lorentz invariants $s_{ij}:=2k_i\cdot k_j$ and the moduli space of Riemann spheres with $n$ marked points $\mathfrak{M}_{0,n}$ \cite{Fairlie-Roberts-1972, Gross-Mende-1987, Witten:2004cp, Cachazo:2013iaa, Fairlie:2008dg, Cachazo:2013gna},
\begin{align}\label{SE-CHY}
  f_a \,=\, \sum_{b\neq a} {s_{ab} \over z_a  - z_b} \,=\, 0,
  \quad a=1,2, \ldots, n,
\end{align}
where the unknowns $z_a \in \mathbb{CP}^1$ denote punctures on the Riemann sphere, and the kinematical invariants $s_{ab}$ satisfy momentum conservation and on-shell conditions, i.e.~$\sum_{b\ne a}s_{ab} = -s_{aa} = 0$.
This system has a global $\operatorname{SL}(2,\mathbb{C})$ symmetry, and thus only $n{-}3$ out of the $n$ equations are independent.
It has been proven that the number of independent solutions to the scattering equations is $(n{-}3)!$ \cite{Cachazo:2013gna, Dolan:2014ega}.

In a new formalism developed by Cachazo, He and Yuan (CHY) \cite{Cachazo:2013hca, Cachazo:2013iea, Cachazo:2014nsa, Cachazo:2014xea}, the tree-level S-matrix in massless field theories is expressed as a multiple integral over the $\mathfrak{M}_{0,n}$.
The integral is fully localized to the zeroes of the scattering equations and can be written as a sum over residues
\begin{align}\label{CHY-sum-sols}
  {\cal A}_n  \,=\, \sum_{i=1}^{(n-3)!} {{\cal I}_n(z,k) \over {\det}'\Phi(z,k)}\bigg|_{z=z^{(i)}},
\end{align}
where $z^{(i)} \equiv (z_1^{(i)}, \ldots, z_{n}^{(i)})$ stands for the $i^\text{th}$ solution and ${\det}'\Phi$ is the relevant Jacobian determinant (see~e.g.\,\cite{Cachazo:2013hca, Cachazo:2012da} for its explicit expression).
The scattering equations are theory-independent, while the function ${\cal I}_n$ encodes dynamics of the specific theory.
We do not show the precise form of $\mathcal{I}_n$ for any theory, since this paper focuses mainly on the scattering equations.

By now various aspects of the scattering equations and the CHY formalism have been investigated.
In the framework of the scattering equations, new representations for scattering amplitudes in a variety of theories (see e.g.\,\cite{Cachazo:2013hca, Cachazo:2013iea,Cachazo:2014nsa,Cachazo:2014xea,He:2016iqi,Azevedo:2017lkz,Cachazo:2016njl,Heydeman:2017yww,Cachazo:2018hqa}), as well as for form factors in gauge theory \cite{He:2016dol,He:2016jdg,Brandhuber:2016xue}, have been proposed at the tree level.
These new formulas have also shown the power to reveal new mathematical structures behind amplitudes, for example the derivation of various soft theorems \cite{Schwab:2014xua,Afkhami-Jeddi:2014fia,Kalousios:2014uva,Zlotnikov:2014sva,Cachazo:2015ksa,Zlotnikov:2017ahq,Chakrabarti:2017zmh}.
The concrete connection between the CHY formalism and the ambitwistor string theory has been established \cite{Mason:2013sva,Casali:2015vta}.
In particular, the ambitwistor string formalism provides a systematic approach to extend the CHY formulation beyond the tree level \cite{Adamo:2013tsa, Geyer:2015bja,Geyer:2015jch,Geyer:2016wjx,Geyer:2017ela,Geyer:2018xwu,Cachazo:2015aol}.

The scattering equations are universal for all theories, and are fundamental objects in quantum field theory as well as string theory.
They shed new light on the perturbative S-matrix and even the theory itself.
Due to the universality of the scattering equations, the CHY formalism provides an elegant representation for exposing relations between different theories \cite{Cachazo:2013hca,Cachazo:2013iea,Cachazo:2014xea}, such as the famous double-copy relation between amplitudes in Yang-Mills theory and Einstein gravity \cite{Kawai:1985xq, Bern:1998sv, Bern:2008qj}.
Very recently, it was observed that the scattering equations can be interpreted geometrically as a diffeomorphism from the worldsheet associahedron to the kinematic associahedron \cite{Arkani-Hamed:2017mur}.
This gives new insight into the origin of the scattering equations and the CHY formalism \cite{He:Amps2018-talk}.

Due to the importance, it is crucially important to solve the scattering equations.
Notwithstanding efforts have been made to solve the scattering equations or evaluate the CHY formulas \cite{Cachazo:2013gna, Dolan:2014ega, Kalousios:2013eca, Weinzierl:2014vwa, Cardona:2015ouc, Kalousios:2015fya, Sogaard:2015dba, Lam:2015sqb, Bosma:2016ttj, Zlotnikov:2016wtk, Farrow:2018cqi, Huang:2015yka, Dolan:2015iln}, a good method is still missing.
In this paper, we close this gap:~we develop an efficient technique to solve the scattering equations based on the numerical algebraic geometry.

\vskip 5pt
\noindent{\sffamily\bfseries Homotopy continuation~}
Let us give a brief introduction to homotopy continuation \cite{Allgower:1990, Li:1997}, which is the primary method in numerical algebraic geometry that we will use throughout this paper.
In order to solve a system of equations $p(z) = 0$ with $p \equiv(p_1, \ldots, p_N)$ and $z \equiv(z_1, \ldots, z_N)$, the basic idea is to introduce a continuous deformation ({\it homotopy}) $p(z)\to p(z,t)$, $t\in[0,1]$, that connects the target system $p(z,1)=p(z)$ with a start system $p(z,0)=q(z)$ whose solutions $z(0)$ are known.
Then the solutions $z(1)$ of the target system can be obtained from $z(0)$ via smooth paths as the continuation parameter $t$ varies from $0$ to $1$.
To be explicit, constructing a differentiable homotopy $p(z,t)$ and differentiating it with respect to $t$ lead to a system of {\it ordinary differential equations (ODEs)} on $z=z(t)$ as follows:
\begin{align}\label{ODEs}
  {\ud p_i(z,t) \over \ud t} \,=\, \sum_{j=1}^N {\partial p_i(z,t) \over \partial z_j}\, {\ud z_j(t) \over \ud t} + {\partial p_i(z,t) \over \partial t} \,=\, 0.
\end{align}
Viewing this as a system of linear equations on $\ud z_i/\ud t$, it can be transformed into the following standard form:
\begin{align}
	\left( \frac{\ud z}{\ud t} \right) \,=\, - \left( \frac{\partial p(z,t)}{\partial z} \right)^{-1} \left(\frac{\partial p(z,t)}{\partial t}\right), \label{ODEs-std}
\end{align}
where terms in parentheses should be understood as matrices.
Providing the initial condition $z(0)$, the desired solutions $z(1)$ of the target system can be obtained by integrating the system of the ODEs \eqref{ODEs-std}.
Usually, numerical algorithms for initial value problems \cite{Stoer-Bulirsch-2002} are applied to obtain an estimate for $z(1)$.
This approximated solution serves as the initial guess of the true solution, and are fed to the Newton method to further improve its precision \cite{Allgower:1990}.

The homotopy continuation method described above has been well-studied, in particular on polynomial systems, during the past decades.
Therefore this technique can be straightforwardly applied to the system of the scattering equations, since it is equivalent to a system of polynomial equations as follows \cite{Dolan:2014ega}:
\begin{align}\label{SE-poly}
  h_m(z)  \,=\, 0, \quad 1 \le m \le n{-}3,
\end{align}
with
\begin{align}\label{SE-poly-def}
  h_m(z)  \,\equiv\, \sum_{I \subset \{2,\ldots,n-1\}, |I| = m}\, \bigg( s_{\{n\}\cup I}\, \prod_{i\in I}z_i \bigg),
\end{align}
where $s_{A} := \sum_{i<j\in A}s_{ij}$, and three punctures have been fixed as $(z_1, z_2, z_{n}) \to (0, 1, \infty)$ by $\text{SL}(2,\mathbb{C})$ invariance.
Following the homotopy continuation method, a frequently used homotopy is: $h_m(z,t) =  t\, h_m(z) + (1{-}t)(z_{m+2}^m{-}1)$.
The advantage of such construction is that the start system has $(n{-}3)!$ known solutions and the number of solutions remains unchanged for any regular $t$.
Although such a homotopy can be used to solve the scattering equations \eqref{SE-poly} in principle, with some experimentations, we found that it is highly inefficient.
One reason is on the technical side, saying that the complexity of evaluating ODEs \eqref{ODEs} corresponding to the polynomial system is too high.
Another reason is that the initial system is significantly different from the target system, thus implies that a lot of steps are spent to reach the target system.

\vskip 5pt
In this paper, we extend the homotopy continuation method to solve the fractional scattering equations \eqref{SE-CHY} by establishing an appropriate homotopy.

Instead of constructing the homotopy for the system of the scattering equations directly, we propose the {\it physical} homotopy in the kinematic space, i.e.~${\cal S} \rightarrow {\cal S}_t$, where ${\cal S}$ is a point in ${\cal K}_n$.
The momentum conservation and on-shell conditions hold for ${\cal S}_t$ at any $t$.
More explicitly, a simple construction is:
\begin{align}\label{Kinematic-homotopy-convex}
    s_{ij}(t) \,=\, (1-t)\,\bar{s}_{ij} + t\,s_{ij},
\end{align}
where $\bar{s}_{ij}$ and $s_{ij}$ are two sets of Mandelstam variables belonging to the physical region of interest in ${\cal K}_n$.
Clearly, as long as on-shell conditions and momentum conservation are satisfied for $\bar{s}_{ij}$ and $s_{ij}$, they are satisfied for $s_{ij}(t)$.
We define the {\it kinematic homotopy}\footnote{Here we abuse terminology a bit.} as a one-parameter smooth path in the kinematic space ${\cal K}_n$, like \eqref{Kinematic-homotopy-convex}.
The physical kinematic homotopy connects different points in ${\cal K}_n$, and this may be used to establish the connection between the physics quantities evaluated at different points.

The kinematic homotopy ${\cal S}_t$ naturally induces a homotopy for the scattering equations
\begin{align}\label{SE-CHY-homotopy}
  f_a(t) \,=\, \sum_{b\neq a} {s_{ab}(t) \over z_a(t) - z_b(t)} \,=\, 0.
\end{align}
Since the physical homotopy preserves on-shellness and momentum conservation, the system has exact $(n{-}3)!$ solutions for any regular $t$.
To proceed, let us use the $\operatorname{SL}(2, {\mathbb C})$ redundancy to fix three punctures, for example $(z_1, z_2, z_{n}) \to (0,1,\infty)$.
The last equation $f_{n}=0$ is then trivially satisfied \cite{Cachazo:2013gna}.
Differentiating other equations with respect to $t$ gives the following system of ODEs:
\begin{align}\label{ODE-SE-chy}
    \sum_{j=3}^{n-1}\Phi_{i j} {\dot z}_j + f_i' \,=\, 0, \quad i\in\{1,2,\ldots, n{-}1\}
\end{align}
with
\begin{align}\label{}
  \dot{z}_i \,\equiv\, {\ud z_i(t) \over \ud t},
  \quad
  \Phi_{ij} \,\equiv\, {\partial f_i(z,t) \over \partial z_j},
  \quad
  f_i' \,\equiv\, {\partial f_i(z,t) \over \partial t}.
\end{align}
A perfect property is that the matrix $\Phi(t)$ has exactly rank $n {-} 3$ at any $t$ \cite{Cachazo:2013gna}.
This ensures that there is no singularity in our algorithm.
To improve numerical stability, we retain all $(n{-}1)$ equations except $f_{n}=0$ which is satisfied trivially, and employ matrix decomposition methods \cite{Stoer-Bulirsch-2002} to generate the standard form, like eq.\,\eqref{ODEs-std}.
Therefore, once the solutions of the scattering equations for $\bar{s}_{ij}$ is known, the solutions for $s_{ij}$ can be obtained by numerically integrating the ODEs.

However, so far the {\it start solutions} (the solutions of the scattering equations for kinematical invariants ${s}_{ij}(0)=\bar{s}_{ij}$) are not readily available yet.
We would like to emphasize that it is highly non-trivial to obtain the start solutions, in particular when the multiplicity $n$ is large.
In order to initiate our program, we develop an algorithm based on the properties of the scattering equations in some special kinematical regions as well as the homotopy continuation technique.
This algorithm will be described in detail in the following.

We employ the homotopy \eqref{Kinematic-homotopy-convex} again, i.e.~$s_{ij}(t) = (1{-}t)\,\hat{s}_{ij} + t \bar{s}_{ij}$.
Here the kinematical invariants $\hat{s}_{ij}$ satisfy
\begin{align}\label{}
    \hat{s}_{1i}>0, \quad \hat{s}_{2i}>0, \quad \hat{s}_{ij}>0, \quad  i,j \in \{3, \ldots, n{-}1\},
\end{align}
which are referred to as the {\it positive region} denoted by ${\cal K}_n^+$ in \cite{Cachazo:2016ror}.
A remarkable property is that all $(n{-}3)!$ solutions of the scattering equations in ${\cal K}_n^+$ are real \cite{Cachazo:2016ror}.
More interestingly, after using the gauge fixing condition given previously, all puncturs $(z_3, \ldots, z_{n-1})$ live inside the interval $(0,1)$ and  distinct from each other for each solution.
It is clear that due to this feature, the scattering equations in ${\cal K}^+_n$ can be solved much more easily, compared to generic kinematic regions.
As will be detailed below, all $(n{-}3)!$ real solutions can be obtained using the homotopy continuation technique too.\footnote{In \cite{Cachazo:2016ror}, for the kinematics in the positive region, one kind of algorithms were proposed based on interpreting the scattering equations as the equilibrium equations for a stable system of $n{-}3$ particles on the real interval $(0,1)$.}
Once these solutions are readily available, they will serve as start solutions, and we can use the homotopy \eqref{Kinematic-homotopy-convex} and integrate the system of the ODEs \eqref{ODE-SE-chy} to generate the solutions for general kinematics $\bar{s}_{ij}$. 
It is also worth stressing that we can encounter singularities if we still adopt the real contour for $t$ from $0$ to $1$, since the starting and target points live in unphysical and physical regions of ${\cal K}_n$ respectively.
A solution to avoiding the singularities is to employ a complex contour for $t$.
In our program, we choose a simple contour consisting of two line segments in the complex $t$ plane: $0\to 0.5+0.5\,i\to1$.

Now the final task is to obtain all solutions to the scattering equations for one point in ${\cal K}_n^+$.
Inspired by the soft limit of the scattering equations, we propose the following homotopy\footnote{Inspired by the soft limit, one alternative algorithm was constructed and implemented in {\tt Mathematica} in \cite{Cachazo:2013gna}.}
\begin{align}\label{soft-homotopy}
\begin{aligned}
  &\hat{s}_{1i}(t) \,=\, \hat{s}_{1i}, ~~
  \hat{s}_{2i}(t) \,=\, \hat{s}_{2i}, ~~~~3\le i \le n{-}2
  \\
  &\hat{s}_{ij}(t) \,=\, \hat{s}_{ij}, ~~~~3\le i<j \le n{-}2
  \\
  &\hat{s}_{a, n-1}(t) \,=\, t\,\hat{s}_{a, n-1}, ~~~~ 1 \le a \le n{-}2.
\end{aligned}
\end{align}
All the remaining kinematic invariants can be easily obtained via on-shell conditions and momentum conservation.
Clearly, this homotopy preserves the ``positivity'' of the kinematic region ${\cal K}_n^+$.
Another remarkable property is that, in the limit $t\to 0$ which defines the soft limit $k_{n-1} \to 0$, the kinematic space of $n$ particles is reduced to $(n{-}1)$-particle one which is still in positive region.
In this limit, $f_{n-1}(t)$ is invariant up to a factor $t$, i.e.,
\begin{align}\label{soft-SE}
  f_{n-1}(t) = t\,\tilde{f}_{n-1}(t),
  \quad
  \tilde{f}_{n-1}(t) \,=\, \sum_{a=1}^{n-2} {{\hat s}_{a,n-1} \over z_{n-1} {-} z_a},
\end{align}
while other equations become nothing but the system of scattering equations associated with $(n{-}1)$ particles without the soft leg in ${\cal K}_{n-1}^+$.

In order to solve the scattering equations in ${\cal K}_{n-1}^+$, we can use the {\it inverse soft homotopy} \eqref{soft-homotopy} recursively until the four-particle case, whose unique solution is known, i.e.\,$z_3 = {-s_{12}/s_{13}}$ with gauge fixing $(z_1, z_2, z_4) \to (0, 1, \infty)$.
The equation corresponding to the soft particle $\tilde{f}_{n-1}=0$ (referred to as the {\it soft equation}) is equivalent to a polynomial equation of degree $n{-}3$ in $z_{n-1}$.
For each solution of the scattering equations for the $(n{-}1)$-point system without the soft particle,
the $n{-}3$ zeroes of the soft equation $\tilde{f}_{n-1} (z_{n-1}) = 0$ are distributed in the $n{-}3$ sub-intervals of $(0,1)$, separated by $z_3$, $z_4$, $\cdots$, $z_{n-2}$.
Thus simple numerical techniques such as the bisection method can be applied to obtain all $n{-}3$ roots.
For using the inverse soft homotopy \eqref{soft-homotopy} each time, the similar method can be used to solve the soft equation.
Here it should be noted that the $f_{n-1}(t)$ is always replaced by $\tilde{f}_{n-1}(t)$ when we employ the inverse soft homotopy \eqref{soft-homotopy}.
Finally, we can obtain all $(n{-}3)!$ solutions to the scattering equations for one point in ${\cal K}_n^+$.

With the start solutions from solving the scattering equations in ${\cal K}_n^+$, by integrating the corresponding differential equations given in \eqref{ODE-SE-chy}, we can obtain the solutions to the scattering equations for one point in ${\cal K}_n$.

\vskip 3pt
To summarise, we have proposed a homotopy continuation method to solve the scattering equations and given a workable framework in detail.
As shown schematically below (superscript (s) stands for the soft limit), our method consist of two main steps.
\begin{align}\label{SE-homotopy}
\begin{aligned}
    \underbrace{
    {\cal K}_5^{+(\text{s})}
    \xlongrightarrow{\eqref{soft-homotopy}}
    \cdots
    \xlongrightarrow{\eqref{soft-homotopy}}
    {\cal K}_n^{+(\text{s})}
    \xlongrightarrow{\eqref{soft-homotopy}}
    {\cal K}_n^{+}
    \xlongrightarrow{\eqref{Kinematic-homotopy-convex}}
    \hspace{5pt}
    }_\text{Step I}
    \hspace{-5pt}
    {\cal K}_n
    \hspace{-10pt}
    \underbrace{
    \hspace{9pt}
    \xlongrightarrow{\eqref{Kinematic-homotopy-convex}}
    {\cal K}_n
    }_\text{Step II}
\end{aligned}
\end{align}
The first step is to obtain the start solutions, which consists of two substeps:
First, solve the scattering equations in ${\cal K}_n^+$ by using the inverse soft homotopy \eqref{soft-homotopy} recursively.
Then, with these solutions as start solutions, we can use the homotopy \eqref{Kinematic-homotopy-convex} to solve the scattering equations for one point in the realistic target region.
As the next step, once we have all $(n{-}3)!$ solutions to the scattering equations for one physically realistic point in the kinematic space, we can track these solutions to any point in the kinematic space using the homotopy \eqref{Kinematic-homotopy-convex}.
In the second step, the solutions of the start system can be continued to the target system much more easily, since they both live in the same physically realistic region.

\vskip 5pt
The method presented above has been implemented  into a {\tt C++} program.
For the numerical integration of differential equations, we adopt the Runge-Kutta-Fehlberg 7(8)-th order method \cite{Fehlberg:1968} provided by \textsc{Odeint} \cite{odeint}; and for the numerical solution of linear equation system, we adopt the Householder QR decomposition with column pivoting provided by \textsc{Eigen} \cite{eigenweb}.
In obtaining the start solutions, the local accuracy is set to be $10^{-15}$, while in the second step, the local accuracy is set to be $10^{-7}$.
In both steps, the Newton method is adopted to increase the precision to $10^{-15}$.
The code is available at [\href{https://github.com/zxrlha/sehomo}{\tt https://github.com/zxrlha/sehomo}].

We consider the randomly selected non-exceptional points in the phase space corresponding to $2\to n{-}2$ scattering up to $n=13$.
All tests were performed on a Macintosh laptop with a 2.7 GHz processor.
The results of the computation times are summarized in Table \ref{tab-result}.
In the table, $t_n$ are the computation times for obtaining all $\sharp(n) = (n{-}3)!$ solutions, and $\bar{t}_n \equiv t_n/(n{-}3)!$ represents the average time for each solution, for solving the scattering equations with a set of prepared initial solutions in the physically realistic region of ${\cal K}_n$.
That is to say, they correspond to the Step II shown in \eqref{SE-homotopy}.
Here we would also like to note that our algorithm for obtaining the start solutions (i.e.~the Step I in \eqref{SE-homotopy}) works well.
In this step, the time cost is dominated by tracking solutions from unphysical positive region to physically realistic region in ${\cal K}_n$, while solving the scattering equations in the positive region recursively is very fast.
For example, it costs less than 30 minutes for $n=11$ case.
\begin{table}[h]
\begin{center}
\begin{tabular}{c r c c}
  \hline\hline
  ~~$n$~~ & $\sharp(n)$~~ & $~~~~~t_n$~~~~~  &  $~~\bar{t}_n$\,(ms)~~ \\
  \hline
  5   & 2~~                   & $1.3\,\mathrm{ms}$\,~      & 0.7  \\ \hline
  6   & 6~~                   & $5.0\,\mathrm{ms}$\,~     & 0.8 \\ \hline
  7   & 24~~                 & $35\,\mathrm{ms}$~      & 1.5 \\ \hline
  8   & 120~~               & $0.22\,\mathrm{s}$\,~~~~~    & 1.8 \\ \hline
  9   & 720~~               & $1.3\,\mathrm{s}$~~~~  & 1.8 \\ \hline
  10 & 5040~~             & $13\,\mathrm{s}$~~~ & 2.5 \\ \hline
  11 & 40\,320~~         & $2.3\,\mathrm{min}$ & 3.2 \\ \hline
  12 & 362\,880~~       & $\,30\,\mathrm{min}$ & 4.9 \\ \hline
  13 & 3\,628\,800~~   & $5.6\,\mathrm{h}$\,~~~ & 5.5 \\ \hline
  \hline
\end{tabular}
\end{center}
\caption{The total time costs $t_n$ of solving $n$-point scattering equations are shown. The number of solutions as well as the averaged time per solution $\bar t_n\equiv t_n/(n-3)!$ are also shown.\label{tab-result}}
\end{table}

\vskip -10pt
As a consequence of the Newton method, all solutions can be obtained with an accuracy of $10^{-15}$.
We have also checked that all solutions are distinct each other, thus we can verify that no solution is missed.

We observed that the total time to obtain all solutions increases significantly as $n$ increases, mainly due to a factorial increase in the number of solutions.
On the other hand, the average time of obtaining one solution increases much more slowly, and it is still at $\mathcal{O}({\rm ms})$ level even for $n=13$.
It is noteworthy that obtaining different solutions are completely independent, thus can be done in parallel.

We also found that the time costs are dominated by solving the differential equations.
Therefore if higher precision on solutions are requested, only the last step, i.e.~the Newton iterations should be performed within higher precision, which have only small impact on the total time cost.

In addition, due to the property of the algorithm, for two neighboring points in the phase space, clearly it will be much easier to obtain the solutions of the scattering equations from each other.
Therefore, one could speed up the calculation through a book-keeping method:~first the initial solutions are prepared at several typical kinematic points rather than only one point, and the closest point are adopted as the initial point when do actual calculation.

Lastly, let us make a comparison between methods in our paper and in ref.~\cite{Farrow:2018cqi}.
In four dimensions in the spinor-helicity formalism, the scattering equations can be decomposed into `helicity sectors' and written in terms of  two-component spinors with additional variables involved (see~e.g.~refs.~\cite{Roiban:2004yf, Geyer:2014fka, He:2016vfi,Duhr:2018ppq}).
In ref.~\cite{Farrow:2018cqi}, a method was introduced to solve the spinor-valued scattering equations proposed in ref.~\cite{Geyer:2014fka} and implemented in {\tt Mathematica}.
Overall, our algorithm is much faster than the one in ref.~\cite{Farrow:2018cqi} for obtaining all $(n{-}3)!$ solutions.
Here we identify some significant differences as follows.
As already pointed out, obtaining solutions is completely independent of each other in our algorithm.
In contrast, in ref.~\cite{Farrow:2018cqi} the solutions are obtained sequentially, and as more solutions obtained, finding the next solution becomes increasingly difficult. 
Consequently, we can easily obtain all solutions for high points (e.g.~$n=13$), while even for $n=10$ it is quite challenging to solve the equations for all helicity sectors by the package in ref.~\cite{Farrow:2018cqi}.

\vskip 5pt
\noindent{\sffamily\bfseries Conclusion and outlook~}
In this paper we have proposed the kinematic homotopy which connects different points in kinematic space.
Such a homotopy always preserves momentum conservation and on-shell conditions.
With the physical homotopy, we developed an efficient algorithm to generate all numerical solutions of the scattering equations.
This opens a new window of opportunity for further explorations in various prospectives.

First of all, this powerful method allows us to solve the scattering equations with high accuracy and high efficiency in different contexts.
It is interesting to investigate the properties of the scattering equations and the CHY formulas in various kinematical regions, such as collinear and multi-Regge limits.
While the discussion above is limed at the tree level, our method can be simply generalized to solve the scattering equations at loop level, which have been derived from ambitwistor strings.

In practical terms, it allows one to develop a new framework to compute scattering amplitudes at tree and loop level.
Once one obtains all solutions to the scattering equations, as a next step, it is straightforward to generate tree amplitudes or loop integrands by summing up the contributions from these solutions.
For instance, since the scheme to extend the CHY formalism to loop level has been developed at least for gauge and gravity theories, this makes possible to compute the amplitudes in these theories up to the two loop order.

More interestingly, the kinematic homotopy developed in this paper has further significance beyond solving the scattering equations.
It is intriguing that the kinematic homotopy may provide an avenue to study various physical quantities, such as scattering amplitudes and scattering forms \cite{Arkani-Hamed:2017mur, Arkani-Hamed:2017tmz, He:2018okq}, in the kinematic space directly.

\vskip 5pt
\noindent{\sffamily\bfseries Acknowledgments~}
We are grateful to Claude Duhr, Song He, Fabio Maltoni, Jun-Bao Wu and Ellis Yuan for useful comments on the manuscript.
We would also like to acknowledge the hospitality of Galileo Galilei Institute in Florence, ZL also acknowledges the hospitality of the CERN theory division in Geneva and ITP, CAS in Beijing.
ZL is particularly grateful to Hongbao Zhang for his kind hospitality and generous support during a visit to Beijing Normal University.
The work of ZL was supported by the ``Fonds Sp\'{e}cial de Recherche'' (FSR) of the UCLouvain.
This work has received funding from the European Union's Horizon 2020 research and innovation programme as part of the Marie Sk\l{}odowska-Curie Innovative Training Network MCnetITN3 (grant agreement no.\,722104).

\end{document}